\documentclass{svproc}
\usepackage{makecell}
\usepackage{graphicx}
\usepackage{cite}
\usepackage{amsmath,amssymb,amsfonts}
\usepackage{algorithmic}
\usepackage{graphicx}
\usepackage{textcomp}
\usepackage{xcolor}
\usepackage{comment}
\usepackage{hyphenat}
\usepackage{multirow}
\usepackage{tabu}
\usepackage{caption}
\usepackage{subcaption}
\usepackage{graphicx}
\usepackage{caption}
\usepackage{subcaption}
\usepackage{pgfplots}
\usepackage{comment}

\usepackage[numbers,sort&compress,square]{natbib}
\usepackage{subcaption}
\usepackage{
tikz,
relsize,
amsmath,
booktabs,
tikz
}
\usetikzlibrary{arrows,calc,positioning,shadows,shapes}
\usepackage{pgfplots}
\usepackage{pgfplotstable}
\pgfplotsset{compat=1.6}

\usepackage[T1]{fontenc}
\usepackage[utf8]{inputenc}
\usepackage{lmodern}

\usepackage[
labelfont=sf,
hypcap=false,
format=hang,
width=0.8\columnwidth
]{caption}

\usetikzlibrary{
calc,trees,shadows,positioning,arrows,chains,
decorations.pathreplacing,
decorations.pathmorphing,
decorations.shapes,
decorations.text,
shapes,
shapes.geometric,
shapes.symbols,
matrix,
patterns,
intersections,
fit}

\pgfdeclarelayer{background layer}
\pgfdeclarelayer{foreground layer}
\pgfsetlayers{background layer,main,foreground layer}

\tikzset{
>=latex
}
\usepackage[scr=rsfs]{mathalpha}
\usepackage{amsmath}
\usepackage{amssymb}
\usepackage{arydshln}
\usepackage{mathrsfs}
\usepackage[export]{adjustbox}
\usepackage{float}

\usepackage{times}
\usepackage{fancyhdr,graphicx,amsmath,amssymb}
\usepackage[ruled,vlined]{algorithm2e}
\include{pythonlisting}

\def\BibTeX{{\rm B\kern-.05em{\sc i\kern-.025em b}\kern-.08em
    T\kern-.1667em\lower.7ex\hbox{E}\kern-.125emX}}

\usepackage[T1]{fontenc}
\usepackage{graphicx}
\usepackage{tikz}

\begin{document}
\mainmatter 

\title{Direct Comparative Analysis of Nature-inspired Optimization Algorithms on Community Detection Problem in Social Networks}

\titlerunning{Direct Comparative Analysis of Nature-inspired Optimization Algorithms on Community Detection Problem}  % abbreviated title (for running head)
\author{Soumita Das\inst{1} \and Bijita Singha\inst{1} \and Alberto Tonda\inst{2}\and Anupam Biswas\inst{1}
}
\authorrunning{Das et. al} % abbreviated author list (for running head)
\tocauthor{Soumita Das, Anupam Biswas}
\institute{Department of Computer Science and Engineering,\\
National Institute of Technology, Silchar-788010, Assam, India  \\
\and
 Université Paris-Saclay, INRAE,\\
UMR 518 MIA, Palaiseau, France\\
Email:\email{wingsoffire72@gmail.com,bijitasingha7@gmail.com, alberto.tonda@inrae.fr, anupam@cse.nits.ac.in}
}

\maketitle              % typeset the header of the contribution
\begin{abstract}
Nature-inspired optimization Algorithms (NIOAs) are nowadays a popular choice for community detection in social networks. Community detection problem in social network is treated as optimization problem, where the objective is to either maximize the connection within the community or minimize connections between the communities. To apply NIOAs, either of the two, or both objectives are explored. Since NIOAs mostly exploit randomness in their strategies, it is necessary to analyze their performance for specific applications. In this paper, NIOAs are analyzed on the community detection problem. A direct comparison approach is followed to perform pairwise comparison of NIOAs. The performance is measured in terms of five scores designed based on prasatul matrix and also with average isolability. Three widely used real-world social networks and four NIOAs are considered for analyzing the quality of communities generated by NIOAs.

\end{abstract}
\section{Introduction}
%%%%%%%%%%%%%%%%%%%%%%%%%%%%%%%%%%%%%%%%%%%%%%%%%%%%%%%%%%%%%%%

%The main characterizing features of a good nature-inspired algorithm are a higher convergence rate, less processing time, an unbiased exploration and exploitation and less number of algorithmspecific control parameters. Convergence rate is the speed in terms of the number of iterations beyond which a repeated sequence is produced by an algorithm. This repeated sequence is known as a convergent sequence and is closer to the desired solution. The time required for execution of an algorithm is known as processing time. Exploration and exploitation are the elementary parameters of any nature-inspired algorithm. In exploration, exclusive new region of a search space is visited. In exploitation, only the neighborhood region of previously visited points of a search space is visited [36]. A good algorithm requires equilibrium between exploration and exploitation.
%optimization?
% few of the several nature inspired algorithms proposed till now, have  proved to be very efficient. Many algorithms give adequate results, but no algorithm gives an admirable performance in solving of all the optimization problems. In other words, an algorithm may show good performance for some problems while it may perform poorly for other problems [219]. However, as compared to classical optimization techniques, nature-inspired algorithms obtain optimal solutions for a wider range of problem domains in a reasonably practical time
%Here, the value of the dimension is equal to the number of network nodes.

In today's world, majority of the problems are complex in nature and requires optimization of diverse objectives such as minimization of costs, energy consumption and/ or maximization of efficiency, sustainability and performance. Specifically, optimization problems are often subject to a set of complex, non-linear constraints. To solve optimization problems in an effective and time efficient manner, numerous Nature-inspired Optimization Algorithms (NIOAs) are developed~\cite{biswas2013physics,talbi2009metaheuristics,nadimi2021dmfo}. NIOAs are typically based on randomization concept and are used for both continuous and discrete optimization problems. An extensive comparative study of several NIOAs algorithms for continuous and discrete optimization has been performed in~\cite{elbeltagi2005comparison,sarkar2022comparative}. In another work~\cite{sureja2012new}, a comparative analysis of NIOAs on ten continuous and discrete optimization problems has been carried out. In addition to this, numerous methods have been introduced which developed the discrete version of a continuous optimization problem~\cite{taghian2018comparative,biswas2017regression}. An example of a discrete optimization problem is community detection. It is discrete in the sense that each of the solution element in a solution vector with N-dimensions can take only discrete values.  Several NIOAs algorithms on community detection have been proposed~\cite{liu2016community}. Comparative study of few NIOA based community detection has also been carried out~\cite{biswas2015empirical,biswas2017analyzing}

The general principle to solve the community detection problem is to maximize intra-community connectivity (vertices/ entities of the same community are strongly connected) and minimize inter-community connectivity (vertices/ entities belonging to different communities are loosely connected). However, the measure of cohesiveness may vary depending on the type of network (unweighted, weighted, directed, undirected, multiple edges, dynamic etc.). In this paper, we have considered only undirected and unweighted networks for carrying out our experiments and analyze the performance of NIOAs algorithms on community detection. The contributions of this paper are listed as follows:
\begin{itemize}
\item A considerable variety of NIOAs algorithms such as Grey Wolf Optimizer (GWO), Moth-Flame Optimization (MFO), Sine-Cosine Algorithm (SCA) and Whale Optimization Algorithm (WOA) have been used to detect communities in a network. 
\item A comparative performance analysis based on AVerage Isolability (AVI) has been carried to determine the quality of communities identified by the corresponding baselines.

\item Communities obtained from the respective baseline algorithms are directly compared with each other based on D-scores (direct comparison) and K-scores (overall comparison).
\end{itemize}

The organization of the rest of the paper is as follows: Section~\ref{two} emphasizes on the baseline NIOAs algorithms, Section~\ref{three} briefs about the community detection problem, Section~\ref{four} discusses about the direct comparative analysis measure, Section~\ref{five} is dedicated to experimental analysis and Section~\ref{six} concludes the paper.

\begin{figure}
    \centering {\includegraphics[width =10.0cm]{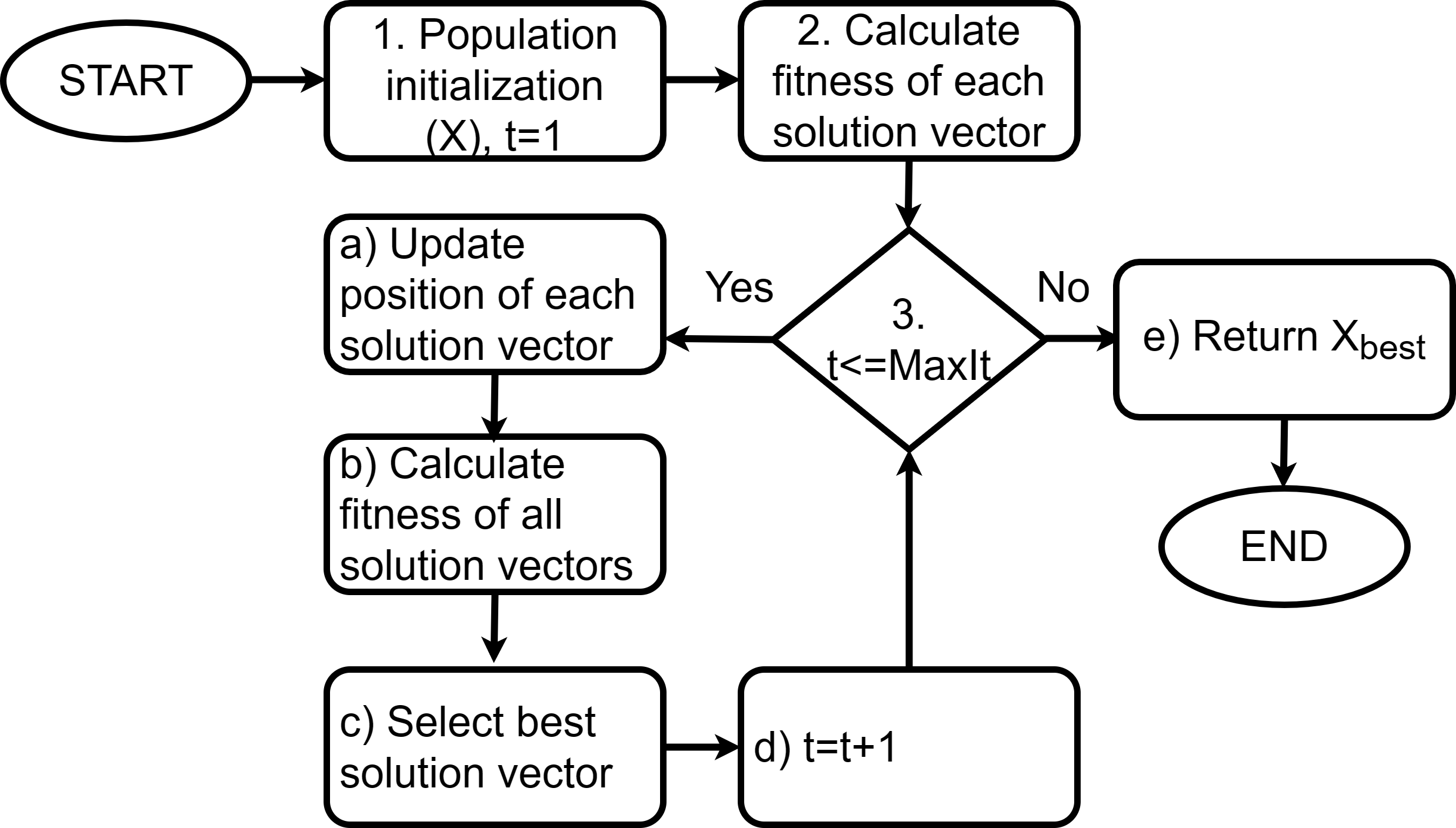}}
    \caption{\small Generic Flow diagram of NIOAs.}
    \label{nioa}
\end{figure}

%%%%%%%%%%%%%%%%%%%%%%%%%%%%%%%%%%%%%%%%%%%%%%%%%%%%%%%%%%%%%%%%%%%%

%As per shown in the below Fig.1 Initialization of population is done using user-defined Initialization function which takes Population size,No.of nodes,and Clusters present in Dataset as parameters.Inside the initialization function each Candidate is being initialized using predefined Random() method that produce real value between[1, clusters] and then Rounded Off this value to its nearest integer using predefined Round() Function. Now two variables Max-iteration is initialized to 500 for optimization purpose and another current-iteration,initially current-iteration=1. For each iteration it is mandatory to check the value of candidate elements whether it lies between [1,clusters] if element value goes lesser than lower bound then updating element value to lower bound that is 1 and if element goes greater than upper bound then updating with upper bound of defined range that is max No.of clusters.Now each candidate is being passed to inner logic program of Nature Inspired Optimization Algorithms and after that current iteration is incremented by 1.Now this process will be executed iteratively till loop condition holds true.After completion of all the iterations, we will get Communities  of the given dataset using Nature Inspired Optimization Algorithms.\\

\section{Nature Inspired Optimization Algorithms}
\label{two}

NIOAs share a set of steps that is portrayed by the generic workflow of the algorithm in Figure~\ref{nioa}. In the first step, the algorithm generates a set of candidate solutions. This candidate solution generation is called population initialization (X) which requires setting of three parameters such as population size, number of dimensions and setting the range of value of solution element. The second step deals with evaluation of the goodness of each of the candidate solution using fitness function. Following this, the termination criteria or the maximum number of iterations (MaxIt) is assigned in third step. Until the termination criteria is satisfied, a set of procedures are repeated as enumerated in the given figure by a), b), c) and d). Firstly, position of each solution vector is updated. Next, the fitness of the updated position vector is computed and compared with the previous fitness. Subsequently, the best solution vector is selected and current iteration counter is incremented by one. Then, after the termination condition is satisfied, the algorithm returns the best solution vector. In this section, we have discussed about some of the best performing algorithms in NIOAs realm which are as follows.

\subsection{Grey Wolf Optimizer (GWO)} This is a population based optimization algorithm inspired by the hunting mechanism of grey wolves found in nature~\cite{mirjalili2014grey}. The wolves are categorized in descending order of leadership hierarchy as $\alpha$, $\beta$, $\delta$ and $\omega$ such that $\alpha$, $\omega$ lies at the top and bottom of hierarchy respectively. GWO algorithm starts with population initialization followed by computation of fitness of wolves where the best three wolves are designated as $\alpha$, $\beta$ and $\delta$. Next, the distance between each wolf and prey is computed by,
\begin{equation}
\overrightarrow{D}=\mid  \overrightarrow{C}.\overrightarrow{X_{p}}(t)-\overrightarrow{X}(t) \mid
\end{equation}
where $t$ represents number of iterations, $\overrightarrow{C}$ indicates coefficient vector, $\overrightarrow{X_{p}}$, $\overrightarrow{X}$ is location vector of prey and grey wolf respectively. Thereafter, position of grey wolf is updated using the following formula,
\begin{equation}
\overrightarrow{X}(t+1)=\overrightarrow{X_{p}}(t)-\overrightarrow{A}.\overrightarrow{D}
\end{equation}
where $A$ is a vector coefficient in [0,2]. Then, position of prey is updated according to the following formula,
\begin{equation}
\overrightarrow{X_{p}}(t+1)=\frac{(\overrightarrow{X_{1}}+\overrightarrow{X_{2}}+\overrightarrow{X_{3}})}{3},
\end{equation}
where $\overrightarrow{X_{1}},\overrightarrow{X_{2}},\overrightarrow{X_{3}}$ represents position vector of $\alpha,\beta,\delta$ wolves respectively. These set of steps are repeated until termination criteria is satisfied. Ultimately, GWO algorithm returns the best position vector for $\alpha$ which indicates the best solution of the problem under consideration.

%It is a population-based meta heuristics algorithm that simulates the leadership hierarchy and hunting mechanism of grey wolves in nature. The Grey Wolf optimizer has been designed to solve single-objective optimization problems.

\subsection{Sine-Cosine Algorithm (SCA)} It is also a population based optimization algorithm where the search for optimal solution is inspired by the sine and cosine trigonometric functions~\cite{mirjalili2016sca}. Initially, SCA algorithm starts with population initialization where each individual is represented by $X_{i}=(x_{i1},...,x{ij},...,x_{iD})$ in the D-dimensional search space. Next, the optimal solution is obtained using  sine and cosine functions depicted by the following formula,

\begin{equation}
\label{sin}
X_{i}^{t+1}=X_{i}^{t}+r_{1}\times \sin(r_{2})\times \mid r_{3}X_{best}^{t}-X_{i}^{t}\mid,~~r_{4} < 0.5
\end{equation}

\begin{equation}
\label{cos}
X_{i}^{t+1}=X_{i}^{t}+r_{1}\times \cos(r_{2})\times \mid r_{3}X_{best}^{t}-X_{i}^{t}\mid,~~r_{4}\ge 0.5,
\end{equation}

where $X_{i}^{t}$ indicates the position of search space at $t^{th}$ iteration, $X_{best}^{t}$ refers to the best position in $t^{th}$ iteration. Equation~\ref{sin} and~\ref{cos} indicates that SCA comprises of four key parameters such as $r_{1}$, $r_{2}$, $r_{3}$ and $r_{4}$ where $r_{1}$ represents the search region. This region lies either between the search agent and target or outside, $r_{2}$ refers to the extent the movement is done towards or outside the target, $r_{3}$ is used to emphasize ($r_{3} > 1$) or de-emphasize ($r_{3} < 1$) the current optimal solution in order to compute the distance to be covered by search agents and $r_{4}$ is used to explore the search space deterministically by switching between sine and cosine functions.

%It obtains a smooth shape for the airfoil with a very low drag, which demonstrates that this algorithm can highly be effective in solving real problems with constrained and unknown search spaces.

\subsection{Moth-Flame Optimization (MFO)}
It is a population based optimization algorithm inspired by the transverse orientation of moths around light sources~\cite{nadimi2021migration}. Moths travel long distances in a straight line by maintaining a fixed angle with the moon. MFO algorithm basically comprises of three primary steps. The first step is population initialization of moths using a matrix $M(t)$ in a D-dimensional search space. Next, fitness of individual moths are stored in an array. 

% Here, moths are the search agents. Suppose, we have $N$ number of moths, then $M(t)$ is represented as,\\
% \begin{equation}
% M(t)= {\begin{bmatrix}
% m_{1,1} & m_{1,2} & \cdots & m_{1,D} \\
% m_{2,1} & m_{2,2} & \cdots  & m_{2,D} \\
% \vdots  & \vdots & \vdots &  \vdots\\
% m_{N,1} & m_{N,2} & \cdots  & m_{N,D} 
% \end{bmatrix} } 
% \end{equation}

% Next, fitness of individual moths are stored in an array as shown below,\\
% \begin{equation}
% OM(t)={\begin{bmatrix}
% OM_{1}(t) \\
% OM_{2}(t) \\
% \vdots \\
% OM_{N}(t) 
% \end{bmatrix}},  
% \end{equation}
% where $t$ determines current number of iteration(s). 
This is followed by storing the flames which are the best positions obtained by moths when searching the search space and is similarly represented in matrix $F(t)$ and it's corresponding fitness values are stored in array $OF(t)$. Next, as the moths come across flames/ artificial light, they try to maintain a similar fixed angle with the flames resulting into a deadly spiral path towards the flames.  Therefore, the second step is associated with updating the position of moths using the following formula,
\begin{equation}
M_{i}(t)=Dis_{i}(t)\times e^{bk} \times \cos(2 \pi k)+F_{j}(t),
\end{equation}

\begin{equation}
Dis_{i}(t)=\mid F_{j}(t)-M_{i}(t) \mid,
\end{equation}
where $M_{i}(t)$ refers to the moth's position in $i^{th}$ iteration, $Dis_{i}(t)$ represents the distance moth $M_{i}(t)$  and corresponding flame $F_{j}(t)$, $k$ is a random number that lies in the range [-1,1], $b$ depicts shape of logarithmic spiral. 

% Then, to converge the algorithm and to provide better exploitation, the number of flames is updated in the third step by using the following equation,\\
% \begin{equation}
% Flame_{Num}(t)=round(N-t\times\frac{N-1}{MaxIt}),
% \end{equation}
% where, $Flame_{Num}(t)$ indicates number of flames at current number, $N$ is total number of flames, $MaxIt$
% is the maximum number of iterations.

\subsection{Whale Optimization Algorithm (WOA)} It is a population based optimization algorithm inspired by the hunting mechanism of humpback whales~\cite{mirjalili2016whale}. Firstly, population of search agents is initialized and fitness of individual search agents is computed. Considering the fitness values, the current best search agent is assumed to be the target prey. Secondly, the position of other search agents are updated near the target prey based on parameters $p$ and $A$. These parameters controls position updating by incorporation of these parameters into three different rules such as encircling prey where  $p<0.5$ and $\mid A \mid < 1$, search for prey where  $p<0.5$ and $\mid A \mid \geq 1$ and spiral updating position where  $p \geq 0.5$. The position of search agent $\overrightarrow{X}(t+1)$ using is updated  by encircling prey at iteration $t+1$ using Equation~\ref{enc1} and Equation~\ref{enc2}.

\begin{equation}
\label{enc1}
\overrightarrow{D}=\mid \overrightarrow{C}. \overrightarrow{X}^{*}-\overrightarrow{X}(t)\mid
\end{equation}

\begin{equation}
\label{enc2}
\overrightarrow{X}(t+1)=\mid \overrightarrow{X}^{*}(t)-\overrightarrow{A}.\overrightarrow{D}\mid,
\end{equation}

\begin{equation}
\overrightarrow{A}=2\overrightarrow{a}.\overrightarrow{r}-\overrightarrow{a}
\end{equation}

\begin{equation}
\overrightarrow{C}=2.\overrightarrow{r}
\end{equation}

 $\overrightarrow{X}^{*}$ represents the  best search agent in the current iteration $t$, $\overrightarrow{X}(t)$ represents position of a search agent at iteration $t$,  the value of $a$ decreases from 2 to 0 over the iterations, $r$ is a random number in range [0,1]. Next, searching for prey is similar to encircling prey. However, the only difference is that $\overrightarrow{X}^{*}$ is replaced with a randomly selected search agent $\overrightarrow{X}_{rand}$. In spiral position update, the positions of individual search agents are updated using the following equation,\\
\begin{equation}
\overrightarrow{X}(t+1)=\overrightarrow{\acute{D}}.e^{bl}.\cos{(2 \pi l)} + \overrightarrow{X^{*}}(t),
\end{equation}
where $\overrightarrow{\acute{D}}=\mid \overrightarrow{X}^{*}-\overrightarrow{X}(t) \mid$ which indicates the difference of the distance between the target prey and the search agent at the current iteration, $b$ is constant, $l \in [-1,1]$. The position of search agents are updated until the termination criteria and finally WOA algorithm returns the best search agent.

\section{Community Detection Problem}
\label{three}

The problem of community detection in networks belong to the class of graph partitioning problem, and it is thus a NP-hard problem~\cite{buluc2013data}. Therefore, it has received a lot of attention in recent years and several community detection methods have been introduced for identifying communities in networks. A network comprises of a set of entities and relationships/ connections shared by the entities. Networks are represented in the form of a graph indicated by $G (V,E)$ comprising of nodes ($V$) referring to entities and edges ($E$) specifying connections. The problem is to divide the network into several communities $C=\{C_{1},C_{2},C_{3},..,C_{k}\}$ where each community say  $C_{i},~~~ \forall i={1,2,..,k}$ consists of a set of nodes belonging to $V$ such that the number of connections within $C_{i}$ should be maximized and number of connections between $C_{i}$ and other communities should be minimized. These maximization or minimization requires the use of fitness function in order to obtain the best solution.

%The fitness of communities are analyzed using community evaluation metrics such as modularity, Normalized Mutual Information (NMI), purity, Adjusted Random Index (ARI) ~\cite{chakraborty2017metrics} etc. Modularity is used to measure the quality of community. Whereas NMI, purity, ARI is used to measure accuracy of community. 

%Moreover, community detection problem is classified as single-objective optimization problem~\cite{} and multi-objective optimization problem~\cite{} based on the cardinality of fitness function used.\\

Suppose, $G(V,E)$ is divided into $l$ feasible partitions $P=\{P_{1},P_{2},P_{3},..,P_{l}\}$. Then, community detection problem is formulated  as an optimization problem using the following equation,\\
\begin{equation}
f(P^{best})=max f(P),
\end{equation}
where $P^{best}$ is the desired partition of the network obtained by incorporating a fitness function $f$ which evaluates the goodness of the network.

\textbf{Fitness function:} It is required to find the best solution in an optimization problem. Here, as we are considering community detection as an optimization problem, so for fitness computation,  community evaluation metrics such as are modularity, Normalized Mutual Information (NMI), purity, Adjusted Random Index (ARI) etc. are used~\cite{chakraborty2017metrics,das2021deployment}. Modularity is used to measure the quality of community, whereas NMI, purity, ARI is used to measure accuracy of community.  Depending on the cardinality of fitness function used, community detection problem is classified as single-objective optimization problem and multi-objective optimization problem~\cite{ferligoj1992direct}.

%single fitness function or multiple fitness function are considered for solving the community detection problem and so the community detection problem is classified as single-objective optimization problem and multi-objective optimization problem~\cite{ferligoj1992direct}.
%Community Detection- The concept of community detection has emerged in network science as a method for finding groups within complex systems through representation on a graph. Detecting communities in a network is one of the important tasks in network analysis.

\section{Direct comparative analysis}
\label{four}
%Particularly, these matrix considers optimality and comparability of solutions obtained from respective algorithms.
The rapid growth of NIOAs have necessitated the performance evaluation of the respective algorithms. Though several statistical measures such as mean, standard deviation and median are used for performance comparison purpose, but these measures do not directly compare the solutions given by two separate algorithms say primary algorithm ($A_{p}$) and alternative algorithms ($A_{q}$), where $A_{p}$ refers to those algorithms whose performance is to be evaluated and $A_{q}$ refers to the set of algorithms with which $A_{p}$ is to be compared. In this paper, we have used D-scores and K-scores for direct comparison and overall comparison respectively  to evaluate the quality of communities~\cite{biswas2022prasatul}. \\

%Prasatul Matrix to compare the solutions given by different algorithms directly. Here, the quality of solution is interpreted based on direct comparison using two scores such as Direct Optimality (DO), Direct Comparability (DC) and  overall comparison using three scores such as Overall Optimality (KO), Overall Comparability (KC) and Overall Together (KT)

\textbf{Direct Optimality (DO:)} $A_{p}$ is compared with $A_{q}$ in terms of optimality by combining the comparative performance considering best performance, average performance and worst performance of $A_{p}$ with respect to $A_{q}$ denoted by $O_{1}, O_{2},O_{3}$ respectively and is defined by,

\begin{equation}
DO=O_{1}+0.5*O_{2}-O_{3}
\end{equation}

\textbf{Direct Comparability (DC):} $A_{p}$ is compared with algorithm $A_{q}$ in terms of three levels of abstractions such as win, tie and loose denoted by $C_{1}$, $C_{2}$ and $C_{3}$ respectively and is defined by,

\begin{equation}
DC=C_{1}+0.5*C_{2}-C_{3}
\end{equation}

\textbf{Overall Optimality (KO):} The overall optimality of $A_{p}$ is computed based on three levels of abstraction such as best, average and worst irrespective of win or loose indicated by $K_{1}^{0}$, $K_{2}^{0}$ and $K_{3}^{0}$ respectively and is defined by,
\begin{equation}
KO=K_{1}^{0}+0.5*K_{2}^{0}-K_{3}^{0}
\end{equation}

\textbf{Overall Comparability (KC):} $A_{p}$ is compared with $A_{q}$ by considering overall comparability in all three levels of abstraction such as win, tie and loose indicated by $K_{1}^{c}$, $K_{2}^{c}$ and $K_{3}^{c}$ respectively and is defined by,
\begin{equation}
KC=K_{1}^{c}+0.5*K_{2}^{c}-K_{3}^{c}
\end{equation}

\textbf{Overall Together (KT):} It is used to interpret that $A_{p}$ performs better than $A_{q}$ considering that abstraction levels such as best $\&$ average and win $\&$ tie are overlapping and is defined by,

\begin{equation}
KT=\frac{a+b+d+e}{n}
\end{equation}
where a,b, c and d represents the overlapping abstraction levels, $n$ indicates total number of possible combinations of abstraction levels.

\begin{table}[t]
\caption {\scriptsize Dataset Statistics. First column contains dataset details, $\# ~Nodes$ refers to number of nodes, $\# ~Edges$ refers to number of edges, Avg. degree indicates average degree of the graph.}

\label{dat}
\centering
 \begin{tabular} {|m{10.2em}|m{6em}|m{6em}|m{7em}|}
 \hline
 \centering{Dataset} & \centering{\# ~Nodes} & \centering{ \# ~Edges} & \centering{Avg. degree} 
 \tabularnewline [4pt]
 %\\
\hline
 \centering{Karate~\cite{karate}} & \centering{34} &  \centering{78}  &  \centering{4.58}  
 \tabularnewline[4pt]
 \centering{Dolphin~\cite{dolphin}} & \centering{62} &  \centering{159}  &  \centering{5.12} 
  \tabularnewline[4pt]
  \centering{Football~\cite{football}} & \centering{115} &  \centering{613}&  \centering{10.66}  
   \tabularnewline [4pt]
   \hline
 % \hline
   \end{tabular}
\end{table}

\section{Experimental Analysis}
\label{five}
%\subsection{Experimental Setup}
In this work, experiments are conducted on several widely used real-world datasets such as karate network~\cite{karate}, dolphin network~\cite{dolphin} and football network~\cite{football} summarized in Table~\ref{dat}. Several state-of-the-art NIOAs algorithms such as GWO, MFO, SCA and WOA have been used on community detection to perform a comparative analysis of these algorithms using average isolability and five different performance measures based on optimality and comparability ~\cite{BISWAS20171}. Also, the performance of NIOAs algorithms for community detection is highly dependent on parameter settings. Therefore, in this section, we discuss about algorithm parameter settings, average isolability and result analysis.

\subsection{Algorithm parameter settings}
There are two types of parameters in NIOAs algorithms namely, common parameters and algorithm specific parameters. Parameters that are common in all NIOAs algorithms are called common parameters and parameters specific to a particular NIOAs algorithm are the algorithm specific parameters. There are particularly three common parameters namely population size, number of dimensions and number of iterations which are described below.

\textbf{Number of dimensions:} In community detection context, number of dimensions is equal to the total number of nodes present in a network. The size of candidate solution is equal to the number of dimensions. Total number of such  candidate solutions indicates population size.

%This value is rounded off to it's nearest integer using Round() function because a node must belong to a discrete community.  Each candidate solution is initialized using Random() method which gives real values in [1,number of communities]. 

\textbf{Population size:}  The population size needs to be carefully initialized because the best solution might be dependent on population size. Setting a high population size improves the search capability but leads to increase in time complexity of the algorithm. In our experiments, we have set the population size as 30.

%textcolor{red}{ First of all, Initialization of population is done using Initialization function which takes Population size, No.of Nodes and Clusters present in Dataset as parameters. Inside the Initialization Function each Candidate is being initialized using Random() method that gives real value between[1,No. of clusters] and then Rounded Off this value to its nearest integer using Round() Function. Current iteration is initialized to 1.\\}

%A population consists of a set of randomly generated candidate solutions. population size refers to the number of randomly generated candidate solutions,
%%%%%%%%%%%%%%%%%%%%%%%%%------------------------------------%%%%%%%%%%%%%%%%

\textbf{Number of iterations:} It is also a key parameter to find the optimal solution. Initially, current iteration is set to 1. For specification of number of iterations, two aspects are to be considered. Firstly, if the number of iterations is small, then the optimal solution might not be found. Whereas, large number of iterations increases time complexity of optimization algorithms and may lead to redundancy i.e. iterations may continue even after attaining the best solution. Therefore, number of iterations must be carefully set. 

%Next, for each iteration it is necessary to check and update the candidate element to boundary value in the range [1,number of communities]. Then, fitness value of each of the candidate solution is obtained followed by execution of inner logic of NIOAs for optimization and after completion of the current iteration, the counter of the current iteration is incremented by 1. This process continues till the specified number of iterations.  Finally, at the end of the maximum number of iterations, communities are obtained. For our experimental purpose, we have set the number of iterations as 500.

%maximized AVerage Isolability (AVI)~\cite{BISWAS20171}. Computation of AVI requires candidate solution, number of nodes, number of clusters and graph representation of dataset as input. It is used to examine the ability of a cluster to isolate itself from rest of the network by examining the nodes based on the strength of connections. 

\subsection{Average Isolability} It is required to compare and improve the candidate solutions to obtain a near optimal solution. In our experiment, we have considered individual cluster specific fitness function namely, AVerage Isolability (AVI)~\cite{BISWAS20171} where the objective is to examine the ability of a cluster to isolate itself from rest of the network by examining the nodes based on the strength of connections. Therefore, to find the optimal solution, we have maximized AVI. For an undirected graph, Isolability of a cluster $C_{i}$ is defined by,

\begin{equation}
\label{iso}    
    \text{Isolability } (C_{i}) = \frac{ \{(u, v) \mid u \in_{C_{i}} v \}}     {\{\{  (u, v); (u, w)\} \mid u \in_{C_{i}} v ~\&  w \notin C_{i}\}}, 
\end{equation}

  where, the numerator term indicates connections within the community $C_{i}$ and denominator is the total number of connections. Next, AVI is defined by,
  \begin{equation}
   Q_{AVI}(G,C)=\frac{1}{K} \sum Isolability (C_i),
\end{equation}
where $k$ indicates total number of clusters in $G(V,E)$.

\begin{table}[t]
\captionsetup{width=\columnwidth}
\caption {\small Comparative performance of MFO algorithm with alternative algorithms based on D-scores and K-scores.}
\label{ds}
\centering
\begin{tabular}{| m{6em}| m{19em} | m{19em}| m{19em}|m{19em}|m{19em}|m{10em}|m{10em}|m{8em}|m{8em}|m{12em}|m{12em}|m{8em}|m{10em}|m{10em}|m{8em}|}
\hline
%\multicolumn{13}{|c|}{Primary Algorithm: MFO} \\ \hline
 %\multicolumn{13}{|c|}{Nodes:~34, Clusters:~2, Dataset:~Karate} \\ \hline
\multirow{2}{*}{\scriptsize{Dataset}}& \multicolumn{5}{c|}{GWO} & \multicolumn{5}{c|}{SCA} & \multicolumn{5}{c|}{WOA} \\ \cline{2-16}
 & \multicolumn{1}{l|}{DO} & \multicolumn{1}{l|}{DC} &\multicolumn{1}{l|}{KO} & \multicolumn{1}{l|}{KC} & \multicolumn{1}{l|}{KT} & \multicolumn{1}{l|}{DO} & \multicolumn{1}{l|}{DC} &\multicolumn{1}{l|}{KO} & \multicolumn{1}{l|}{KC} & \multicolumn{1}{l|}{KT} & \multicolumn{1}{l|}{DO} & \multicolumn{1}{l|}{DC} &\multicolumn{1}{l|}{KO} & \multicolumn{1}{l|}{KC} & \multicolumn{1}{l|}{KT} \\ \hline
\multicolumn{1}{|l|}{Karate} & \multicolumn{1}{l|}{0.98} & \multicolumn{1}{l|}{0.98} &\multicolumn{1}{l|}{1.00} & \multicolumn{1}{l|}{1.00} & \multicolumn{1}{l|}{1.00} & \multicolumn{1}{l|}{0.98} & \multicolumn{1}{l|}{0.98} &\multicolumn{1}{l|}{1.00} & \multicolumn{1}{l|}{1.00} &\multicolumn{1}{l|}{1.00} & \multicolumn{1}{l|}{0.06} & \multicolumn{1}{l|}{0.75} & \multicolumn{1}{l|}{0.53} & \multicolumn{1}{l|}{1.0} & \multicolumn{1}{l|}{1.0} \\ \hline
\multicolumn{1}{|l|}{Dolphin} & \multicolumn{1}{l|}{1.44} & \multicolumn{1}{l|}{0.76} &\multicolumn{1}{l|}{1.00} & \multicolumn{1}{l|}{0.78} & \multicolumn{1}{l|}{1.00} & \multicolumn{1}{l|}{1.44} & \multicolumn{1}{l|}{0.76} &\multicolumn{1}{l|}{1.00} & \multicolumn{1}{l|}{0.78} &\multicolumn{1}{l|}{1.00} & \multicolumn{1}{l|}{0.17} & \multicolumn{1}{l|}{0.17} & \multicolumn{1}{l|}{0.31} & \multicolumn{1}{l|}{0.30} & \multicolumn{1}{l|}{0.64} \\ \hline
\multicolumn{1}{|l|}{Football} & \multicolumn{1}{l|}{1.44} & \multicolumn{1}{l|}{0.75} &\multicolumn{1}{l|}{1.00} & \multicolumn{1}{l|}{0.76} & \multicolumn{1}{l|}{1.00} & \multicolumn{1}{l|}{1.44} & \multicolumn{1}{l|}{0.75} &\multicolumn{1}{l|}{1.00} & \multicolumn{1}{l|}{0.76} &\multicolumn{1}{l|}{1.00} & \multicolumn{1}{l|}{1.44} & \multicolumn{1}{l|}{0.75} & \multicolumn{1}{l|}{1.00} & \multicolumn{1}{l|}{0.76} & \multicolumn{1}{l|}{1.00} \\ \hline
\end{tabular}%
\end{table}

\begin{comment}
\begin{table}[t]
\captionsetup{width=\columnwidth}
\caption {\small Comparative performance of MFO algorithm with alternative algorithms based on D-scores.}
\label{dss}
\centering
\begin{tabular}{| m{32em}| m{14em} | m{14em} | m{14em}|m{14em}|m{14em}|m{14em}|m{10em}|m{14em}|}
 \hline
\multicolumn{9}{|c|}{Primary Algorithm:~MFO} \\ \hline
 \multicolumn{9}{|c|}{Nodes:~34, Clusters:~2, Dataset:~Karate} \\ \hline
\multicolumn{1}{|l|}{\multirow{2}{*}{\scripsize{f}}} & \multicolumn{2}{|c|}{GWO} & \multicolumn{2}{c|}{SCA} & \multicolumn{2}{c|}{WOA} & \multicolumn{2}{c|}{AVG} \\ \cline{2-9} 
\multicolumn{1}{|l|}{} & \multicolumn{1}{l|}{DO} & \multicolumn{1}{l|}{DC} & \multicolumn{1}{l|}{DO} & \multicolumn{1}{l|}{DC} & \multicolumn{1}{l|}{DO} & \multicolumn{1}{l|}{DC} & \multicolumn{1}{l|}{ADO} & ADC \\ \hline
\multicolumn{1}{|l|}{AVI} & \multicolumn{1}{l|}{0.9803} & \multicolumn{1}{l|}{0.9803} & \multicolumn{1}{l|}{0.9803} & \multicolumn{1}{l|}{0.9803} & \multicolumn{1}{l|}{0.0588} & \multicolumn{1}{l|}{0.7500} & \multicolumn{1}{l|}{0.6732} & 0.9035 \\ \hline
\end{tabular}
\end{table}
\end{comment}
  
\subsection{Result Analysis}
Quality of communities given by GWO, MFO, SCA and WOA have been analyzed on three widely used real-world datasets. The analysis has been carried based on the emphasizing on the quality of the  community given by each baseline algorithm and performing comparative evaluation. AVI value is used for quality evaluation. In addition to this, performance analysis based on one-to-one comparison (D-scores) and one-to-many comparison (K-scores) is performed.

%The communities obtained by each of the algorithms on the respective datasets are evaluated based on AVerage Isolability. In addition to this, comparative performance of primary algorithm ($A_{p}$) and alternative algorithms ($A_{q}$) have been analyzed based on D-scores and K-scores.

\pgfplotstableread[row sep=\\,col sep=&]{
AVerage Isolability&GWO   &MFO   &SCA   &WOA\\
Karate          &0.2430   &0.3089   &0.2494   &0.1678\\
Dolphin         &0.1540   &0.1595   &0.1550   &0.1330\\
Football        &0.0288   &0.0329   &0.0289   &0.0294\\
}\lookm

\begin{figure}[!hb]
\centering
\begin{tikzpicture}
    \begin{axis}[
            ybar=0.24cm,
           bar width=.350cm,
            width=0.89\textwidth,
            height=0.50\textwidth,
            enlarge x limits=0.3,
            legend style={at={(0.5,1)},
                anchor=north,legend columns=4,legend cell align=left},
            symbolic x coords={Karate,Dolphin,Football},
            xtick=data,
             x tick label style={rotate=00,anchor=north},
            %nodes near coords,
            nodes near coords align={vertical},
            ymin=0,ymax=0.5,
            ylabel={Average Isolability},
            nodes near coords,
            every node near coord/.append style={rotate=90, anchor=west}
        ]
        \addplot[fill=teal!80, text=black] table[x=AVerage Isolability,y=GWO]{\lookm};
        \addplot[fill=lime!60, text=blue] table[x=AVerage Isolability,y=MFO]{\lookm};
        \addplot[fill=yellow!60, text=black] table[x=AVerage Isolability,y=SCA]{\lookm};
        \addplot[fill=green, text=black] table[x=AVerage Isolability,y=WOA]{\lookm};
        \legend{GWO, MFO, SCA, WOA}
  ,  \end{axis}
\end{tikzpicture}
\caption{\small Comparative analysis of GWO, MFO, SCA and WOA based on Average Isolability.}
\label{AI}
\end{figure}
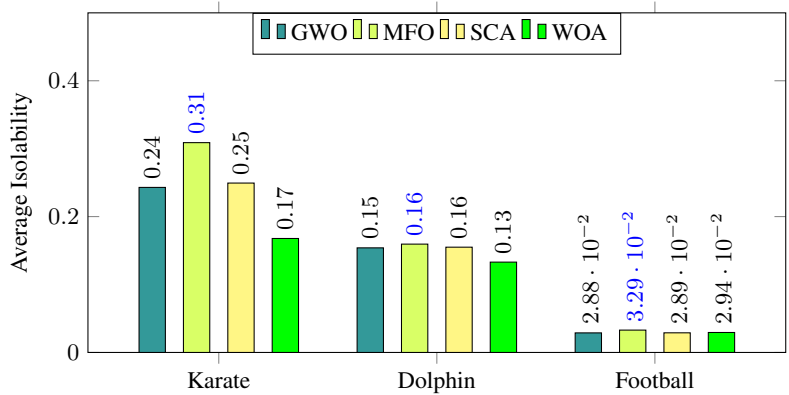

\subsubsection{Result analysis with Average Isolability:}
The AVI scores of the communities given by GWO, MFO, SCA and WOA on real-world datasets are shown in Figure~\ref{AI}. Let us try to analyze the performance of these algorithms with the help of this figure. Here, the X-axis represent real-world datasets namely karate, dolphin and football; Y-axis represents AVI score. The performance of GWO, MFO, SCA and WOA is shown using teal, lime, yellow and green colored bars respectively. The values corresponding to each bar indicates AVI score of the respective algorithms on a given dataset. Higher AVI score indicates good performance of corresponding algorithm and the performance deteriorates with decrease of AVI score. Therefore, the results shown in Figure~\ref{AI} indicates that MFO algorithm gives the best performance on all the datasets and WOA algorithm shows the worst performance on karate and dolphin dataset. Whereas, GWO algorithm shows the worst performance on football dataset.  

% \usepackage{multirow}
% \usepackage{graphicx}

%%%%%%%%%%%%%%%%%%%%%%%%%%%%%%%%%%%%%%%%%%%%%%%%%%%%%%%%%%%%%%%%%%%%%%%%%%%
%%%%%%%%%%%%%%%%%%%%%%%%%%%%%%%%%%%%%%%%%%%%%%%%%%%%%%%%%%%%%%%%%%%%%%%%%%%

%%%%%%%%%%%%%%%%%-------------NEW--------------------

\begin{figure}[!ht]
\centering
%\begin{subfigure}
%\centering
%\newline
\pgfplotstableread[row sep=\\,col sep=&]{
ADO & GWO & MFO & SCA & WOA\\
Karate & 0.0906 & 0.6732 & 0.2176 & 0.9028\\
Dolphin & 0.2482 & 1.0206 & -0.0321 & 1.0795\\
Football & 0.4713 & 1.4430 & 0.2162 & 0.0500\\
}\lookm
\begin{subfigure}{0.56\textwidth}
% \centering
\begin{tikzpicture}[remember picture]
    \begin{axis}[
            ybar=0.1cm,
           bar width=.2cm,
            width=1.0\textwidth,
            height=0.6\textwidth,
            enlarge x limits=0.4,
            legend style={at={(0.49,1.0)},
                anchor=north,legend columns=4,legend cell align=left},
             %legend style={at={(0.5,1)},
             %   anchor=north,legend columns=6,legend cell align=left},
            symbolic x coords={Karate, Dolphin, Football},
            xtick=data,
             x tick label style={rotate=00,anchor=north},
            %nodes near coords,
            nodes near coords align={vertical},
            ymin=-0.3,ymax=2.9,
            ylabel={ADO},
            nodes near coords,
            every node near coord/.append style={font=\tiny,rotate=90, anchor=west}
        ]
        \addplot[fill=teal!80, text=black] table[x=ADO,y=GWO]{\lookm};
       \addplot[fill=lime!60, text=blue]table[x=ADO,y=MFO]{\lookm};
        \addplot[fill=yellow!60, text=black]table[x=ADO,y=SCA]{\lookm};
       \addplot[fill=green, text=black]table[x=ADO,y=WOA]{\lookm};
        \legend{GWO,MFO,SCA,WOA}
         %\legend{DCC,LPA,GM,DER,Gdmp2,Kcut}
    \end{axis}
\end{tikzpicture}

\captionsetup{skip=8pt}
\caption{\scriptsize Comparative analysis based on ADO.}
\label{ris}
\end{subfigure}%
\pgfplotstableread[row sep=\\,col sep=&]{
ADC & GWO & MFO & SCA & WOA\\
Karate & 0.0833 & 0.9036 & 0.0513 & 0.6340\\
Dolphin & 0.1015 & 0.5654 & -0.0287 & 0.8502\\
Football & 0.1555 & 0.7451 & 0.0645 & -0.0190\\
}\lookm
%\begin{subfigure}
%\centering
\begin{subfigure}{0.56\textwidth}
\begin{tikzpicture}[remember picture]
    \begin{axis}[
            ybar=0.1cm,
           bar width=.2cm,
            width=1.0\textwidth,
            height=0.6\textwidth,
            enlarge x limits=0.2,
            %legend style={at={(0.5,1)},
             %   anchor=north,legend columns=6,legend cell align=left},
            symbolic x coords={Karate,Dolphin,Football},
            xtick=data,
             x tick label style={rotate=00,anchor=north},
            %nodes near coords,
            nodes near coords align={vertical},
            ymin=-0.3,ymax=1.2,
            ylabel={ADC},
            nodes near coords,
            every node near coord/.append style={font=\tiny,rotate=90, anchor=west}
        ]
        \addplot[fill=teal!80, text=black] table[x=ADC,y=GWO]{\lookm};
        \addplot[fill=lime!60, text=blue]table[x=ADC,y=MFO]{\lookm};
        \addplot[fill=yellow!60, text=black] table[x=ADC,y=SCA]{\lookm};
      \addplot[fill=green, text=black]table[x=ADC,y=WOA]{\lookm};
         %\legend{DCC,LPA,GM,DER,Gdmp2,Kcut}
    \end{axis}
\end{tikzpicture}
\captionsetup{skip=10pt}
\caption{\scriptsize Comparative analysis based on ADC.}
\label{kar}
\end{subfigure}%
%\caption{Comparative accuracy analysis of edge similarity measures on Strike and Karate dataset}
%\label{figone}
%\end{figure}

%\subsection{Result Analysis:}
%\label{six}

\pgfplotstableread[row sep=\\,col sep=&]{
AKO & GWO & MFO & SCA & WOA\\
Karate & -0.6667 & 0.8433 & -0.6667 & 0.8033\\
Dolphin & -0.7067 & 0.7700 & -0.6267 & 0.7033\\
Football & -0.3600 & 1.000 & -0.3200 & -0.3200\\
}\lookm
%\begin{subfigure}
%\centering
%\newline
\begin{subfigure}{0.56\textwidth}
% \centering
\begin{tikzpicture}[remember picture]
    \begin{axis}[
            ybar=0.1cm,
           bar width=.2cm,
            width=1.0\textwidth,
            height=0.6\textwidth,
            enlarge x limits=0.2,
            %enlarge y limits=-0.6,
                        %legend style={at={(0.5,1)},
             %   anchor=north,legend columns=6,legend cell align=left},
            symbolic x coords={Karate,Dolphin,Football},
            xtick=data,
             x tick label style={rotate=00,anchor=north},
            %nodes near coords,
            nodes near coords align={vertical},
            ymin=-0.9,ymax=2.1,
            ylabel={AKO},
            nodes near coords,
            every node near coord/.append style={font=\tiny,rotate=50, anchor=west}
        ]
        \addplot[fill=teal!80, text=black]table[x=AKO,y=GWO]{\lookm};
       \addplot[fill=lime!60, text=blue]table[x=AKO,y=MFO]{\lookm};
        \addplot[fill=yellow!60, text=black]table[x=AKO,y=SCA]{\lookm};
       \addplot[fill=green, text=black]table[x=AKO,y=WOA]{\lookm}; 
         %\legend{DCC,LPA,GM,DER,Gdmp2,Kcut}
    \end{axis}
\end{tikzpicture}
\captionsetup{skip=8pt}
\caption{\scriptsize Comparative analysis based on AKO.}
\label{ris}
\end{subfigure}%
%%%%
\pgfplotstableread[row sep=\\,col sep=&]{
AKC & GWO & MFO & SCA & WOA\\
Karate & -0.7600 & 1.000 & -0.7567 & 0.9400\\
Dolphin & -0.7467 & 0.6200 & -0.7500 & 0.6600\\
Football & -0.5033 & 0.7600 & -0.4867 & -0.5800\\
}\lookm
%\begin{subfigure}
%\centering
\begin{subfigure}{0.56\textwidth}
% \centering
\begin{tikzpicture}[remember picture]
    \begin{axis}[
            ybar,
           bar width=.2cm,
            width=1.0\textwidth,
            height=0.6\textwidth,
            enlarge x limits=0.4,
            %legend style={at={(0.9,1.7)},
                %anchor=north,legend columns=4,legend cell align=left},
             %legend style={at={(0.5,1)},
             %   anchor=north,legend columns=6,legend cell align=left},
            symbolic x coords={Karate, Dolphin, Football},
            xtick=data,
             x tick label style={rotate=00,anchor=north},
            %nodes near coords,
            nodes near coords align={vertical},
            ymin=-1.0,ymax=2.0,
            ylabel={AKC},
            nodes near coords,
            every node near coord/.append style={font=\tiny,rotate=50, anchor=west}
        ]
        \addplot[fill=teal!80, text=black] table[x=ADO,y=GWO]{\lookm};
       \addplot[fill=lime!60, text=blue]table[x=ADO,y=MFO]{\lookm};
        \addplot[fill=yellow!60, text=black]table[x=ADO,y=SCA]{\lookm};
      \addplot[fill=green, text=black] table[x=ADO,y=WOA]{\lookm};
        %\legend{GWO,MFO,SCA,WOA}
         %\legend{DCC,LPA,GM,DER,Gdmp2,Kcut}
    \end{axis}
\end{tikzpicture}

\captionsetup{skip=8pt}
\caption{\scriptsize Comparative analysis based on AKC.}
\label{ris}
\end{subfigure}%

\pgfplotstableread[row sep=\\,col sep=&]{
AKT & GWO & MFO & SCA & WOA\\
Karate & 0.1400 & 1.000 & 0.1267 & 0.9800\\
Dolphin & 0.1267 & 0.8800 & 0.1133 & 0.8867\\
Football & 0.2600 & 1.000 & 0.2467 & 0.2200\\
}\lookm
%\begin{subfigure}
%\centering
%\newline
\begin{subfigure}{0.56\textwidth}
% \centering
\begin{tikzpicture}[remember picture]
    \begin{axis}[
            ybar,
           bar width=.2cm,
            width=1.0\textwidth,
            height=0.6\textwidth,
            enlarge x limits=0.2,
            %legend style={at={(0.5,1)},
             %   anchor=north,legend columns=6,legend cell align=left},
            symbolic x coords={Karate,Dolphin,Football},
            xtick=data,
             x tick label style={rotate=00,anchor=north},
            %nodes near coords,
            nodes near coords align={vertical},
            ymin=-0.1,ymax=1.8,
            ylabel={AKT},
            nodes near coords,
            every node near coord/.append style={font=\tiny,rotate=90, anchor=west}
        ]
        \addplot[fill=teal!80, text=black] table[x=AKO,y=GWO]{\lookm};
         \addplot[fill=lime!60, text=black]table[x=AKO,y=MFO]{\lookm};
        \addplot [fill=yellow!60, text=black] table[x=AKO,y=SCA]{\lookm};
       \addplot[fill=green, text=black]table[x=AKO,y=WOA]{\lookm}; 
         %\legend{DCC,LPA,GM,DER,Gdmp2,Kcut}
    \end{axis}
\end{tikzpicture}
\captionsetup{skip=8pt}
\caption{\scriptsize Comparative analysis based on AKT.}
\label{ris}
\end{subfigure}%
%%%
\captionsetup{margin=0.2cm}
\caption{Comparative analysis based on average D-score and K-scores i.e. ADO, ADC, AKO, AKC and AKT values for the of communities identified with GWO, MFO, SCA and WOA on real-world datasets.} %on Riskmap and Dolphin dataset}
\label{figthree}
\end{figure}
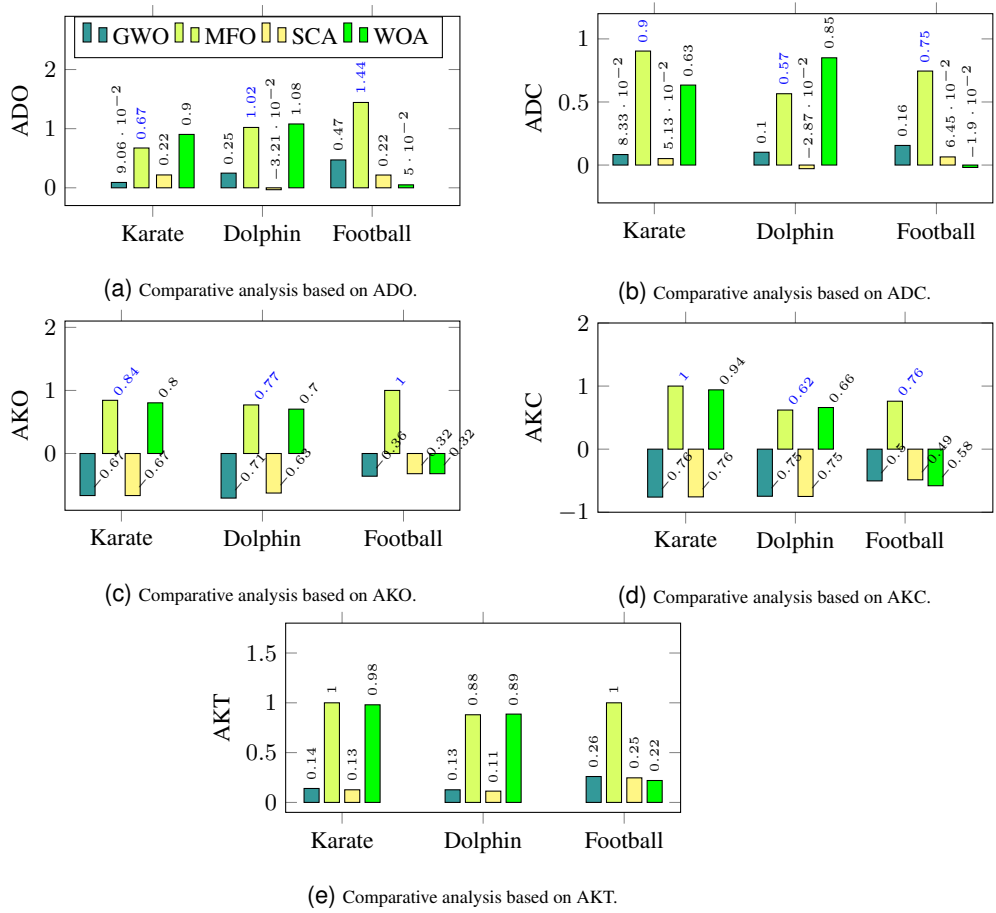

\subsubsection{Result analysis based on D-scores and K-scores:}

D-scores and K-scores are used to evaluate the performance of all possible combinations of the baseline algorithms in terms of the quality of communities given by the respective algorithms. All such combinations of baseline algorithms indicated by $(A_{p},A_{q})$ is considered as a comparable algorithm pair. The results of comparable algorithm pairs such as $(A_{p}= MFO, A_{q}=GWO)$,  $(A_{p}= MFO, A_{q}=SCA)$ and $(A_{p}= MFO, A_{q}=WOA)$  in terms of D-scores and K-scores on karate, dolphin and football dataset are summarized in Table~\ref{ds}. Then, average DO (ADO), average DC (ADC) score, average KO (AKO), average KC (AKC) and average KT (AKT) is obtained by summation of corresponding DO, DC, KO, KC and KT scores of all comparable algorithm pairs with $A_{p}=MFO$ divided by the total number of such pairs and the results are shown in Figure~\ref{figthree}. High ADO, ADC, AKO, AKC, AKT scores indicate that $A_{p}$ performs better than $A_{q}$ in terms of optimality and comparability. For each dataset and corresponding performance measure, highest positive score obtained by the respective algorithm is ranked as 1, second highest is ranked as 2 and so on. Following this ranking procedure, MFO algorithm is ranked as 1 and hence, it is the best performing algorithm in terms of D-scores, K-scores. Following this strategy, SCA gives the worst performance.

    \section{Conclusion}
    \label{six}

A quality measure based on connection strength associated with a cluster called average isolability and a direct comparison approach based on five scores designed based on prasatul matrix is used to evaluate the quality of communities considering optimality and comparability. Four NIOAs and three widely used real-world datasets are used to perform comparative analysis. Results based on average isolability indicate that the MFO algorithm gives the best performance on all datasets. Whereas, WOA algorithm has the worst performance on karate and dolphin datasets, GWO algorithm has the worst performance on football datasets. Following this, the performance analysis based on the five scores derived from prasatul matrix suggests that the MFO algorithm achieves the best performance and the SCA algorithm gives the worst performance.   
    %Community Detection of standard Karate Dataset,Dolphins Dataset and Football Datasets using Nature Inspired Optimization Algorithms(NIOAs) that are inspired by Natural phenomenon(such as GWO, MFO,SCA and WOA) has been performed using Average Isolability as Fitness Function that takes Candidate, No.of Nodes, No. of Clusters and Graph Representation of dataset as parameters. On the basis of Graph it determines the ability of any cluster  to isolate itself from other clusters. Isolability is measured as ratio of connection strength within the cluster to total strength of connections associated with the cluster. After that inner logic program of Nature Inspired Optimization Algorithms has been executed for Optimization and after completion of current iteration counter current\_iteration value value is incremented by 1.\\     This process has been executed iteratively Maximum\_iteration No. of Times.\\    Finally after completion of all the iterations we got Communities of the datasets using Nature Inspired Optimization Algorithms. Interpretation of performance\\    of algorithms are done in  terms of optimality of the solutions and comparability of solutions are determined on the basis of direct comparison and overall comparison,also referred as D-scores and K-scores respectively. After Analysis of D-scores and K-scores Ranking Tables of Karate Dataset, Dolphin Dataset and \\Football Dataset MFO     has been observed as Overall Best Performing Algorithm and SCA is Overall Worst Performing Algorithm in (GWO,MFO,SCA and WOA)Nature Inspired Optimization Algorithms.
\section*{References}
\vspace{-30pt}

\renewcommand{\bibname}{}
%\bibliographystyle{plain}
%\bibliography{mybib}
\let\clearpage\relax
\bibliographystyle{unsrt}
\bibliography{mybib}

\begin{thebibliography}{10}

\bibitem{biswas2013physics}
Anupam Biswas, KK~Mishra, Shailesh Tiwari, and AK~Misra.
\newblock Physics-inspired optimization algorithms: a survey.
\newblock {\em Journal of Optimization}, 2013, 2013.

\bibitem{talbi2009metaheuristics}
El-Ghazali Talbi.
\newblock {\em Metaheuristics: from design to implementation}.
\newblock John Wiley \& Sons, 2009.

\bibitem{nadimi2021dmfo}
Mohammad~H Nadimi-Shahraki, Ebrahim Moeini, Shokooh Taghian, and Seyedali
  Mirjalili.
\newblock Dmfo-cd: A discrete moth-flame optimization algorithm for community
  detection.
\newblock {\em Algorithms}, 14(11):314, 2021.

\bibitem{elbeltagi2005comparison}
Emad Elbeltagi, Tarek Hegazy, and Donald Grierson.
\newblock Comparison among five evolutionary-based optimization algorithms.
\newblock {\em Advanced engineering informatics}, 19(1):43--53, 2005.

\bibitem{sarkar2022comparative}
Debojyoti Sarkar and Anupam Biswas.
\newblock Comparative performance analysis of recent evolutionary algorithms.
\newblock In {\em Evolution in Computational Intelligence}, pages 151--159.
  Springer, 2022.

\bibitem{sureja2012new}
Nitesh Sureja.
\newblock New inspirations in nature: A survey.
\newblock {\em International Journal of Computer Applications \& Information
  Technology}, 1(3):21--24, 2012.

\bibitem{taghian2018comparative}
Shokooh Taghian, Mohammad~H Nadimi-Shahraki, and Hoda Zamani.
\newblock Comparative analysis of transfer function-based binary metaheuristic
  algorithms for feature selection.
\newblock In {\em 2018 International Conference on Artificial Intelligence and
  Data Processing (IDAP)}, pages 1--6. IEEE, 2018.

\bibitem{biswas2017regression}
Anupam Biswas and Bhaskar Biswas.
\newblock Regression line shifting mechanism for analyzing evolutionary
  optimization algorithms.
\newblock {\em Soft Computing}, 21(21):6237--6252, 2017.

\bibitem{liu2016community}
Qiang Liu, Bin Zhou, Shudong Li, Ai-ping Li, Peng Zou, and Yan Jia.
\newblock Community detection utilizing a novel multi-swarm fruit fly
  optimization algorithm with hill-climbing strategy.
\newblock {\em Arabian Journal for Science and Engineering}, 41(3):807--828,
  2016.

\bibitem{biswas2015empirical}
Anupam Biswas, Pawan Gupta, Mradul Modi, and Bhaskar Biswas.
\newblock An empirical study of some particle swarm optimizer variants for
  community detection.
\newblock In {\em Advances in Intelligent Informatics}, pages 511--520.
  Springer, 2015.

\bibitem{biswas2017analyzing}
Anupam Biswas and Bhaskar Biswas.
\newblock Analyzing evolutionary optimization and community detection
  algorithms using regression line dominance.
\newblock {\em Information sciences}, 396:185--201, 2017.

\bibitem{mirjalili2014grey}
Seyedali Mirjalili, Seyed~Mohammad Mirjalili, and Andrew Lewis.
\newblock Grey wolf optimizer.
\newblock {\em Advances in engineering software}, 69:46--61, 2014.

\bibitem{mirjalili2016sca}
Seyedali Mirjalili.
\newblock Sca: a sine cosine algorithm for solving optimization problems.
\newblock {\em Knowledge-based systems}, 96:120--133, 2016.

\bibitem{nadimi2021migration}
Mohammad~H Nadimi-Shahraki, Ali Fatahi, Hoda Zamani, Seyedali Mirjalili, Laith
  Abualigah, and Mohamed Abd~Elaziz.
\newblock Migration-based moth-flame optimization algorithm.
\newblock {\em Processes}, 9(12):2276, 2021.

\bibitem{mirjalili2016whale}
Seyedali Mirjalili and Andrew Lewis.
\newblock The whale optimization algorithm.
\newblock {\em Advances in engineering software}, 95:51--67, 2016.

\bibitem{buluc2013data}
Aydin Buluc, Henning Meyerhenke, Ilya Safro, Peter Sanders, and Christian
  Schulz.
\newblock Recent advances in graph partitioning, 2013.

\bibitem{chakraborty2017metrics}
Tanmoy Chakraborty, Ayushi Dalmia, Animesh Mukherjee, and Niloy Ganguly.
\newblock Metrics for community analysis: A survey.
\newblock {\em ACM Computing Surveys (CSUR)}, 50(4):1--37, August 2017.

\bibitem{das2021deployment}
Soumita Das and Anupam Biswas.
\newblock Deployment of information diffusion for community detection in online
  social networks: a comprehensive review.
\newblock {\em IEEE Transactions on Computational Social Systems},
  8(5):1083--1107, 2021.

\bibitem{ferligoj1992direct}
Anuska Ferligoj and Vladimir Batagelj.
\newblock Direct multicriteria clustering algorithms.
\newblock {\em Journal of classification}, 9(1):43--61, 1992.

\bibitem{biswas2022prasatul}
Anupam Biswas.
\newblock Prasatul matrix: A direct comparison approach for analyzing
  evolutionary optimization algorithms.
\newblock {\em arXiv preprint arXiv:2212.00671}, 2022.

\bibitem{karate}
Wayne~W Zachary.
\newblock An information flow model for conflict and fission in small groups.
\newblock {\em Journal of anthropological research}, 33(4):452--473, 1977.

\bibitem{dolphin}
David Lusseau, Karsten Schneider, Oliver~J Boisseau, Patti Haase, Elisabeth
  Slooten, and Steve~M Dawson.
\newblock The bottlenose dolphin community of doubtful sound features a large
  proportion of long-lasting associations.
\newblock {\em Behavioral Ecology and Sociobiology}, 54(4):396--405, 2003.

\bibitem{football}
Michelle Girvan and Mark~EJ Newman.
\newblock Community structure in social and biological networks.
\newblock {\em Proceedings of the national academy of sciences},
  99(12):7821--7826, 2002.

\bibitem{BISWAS20171}
Anupam Biswas and Bhaskar Biswas.
\newblock Defining quality metrics for graph clustering evaluation.
\newblock {\em Expert Systems with Applications}, 71:1 -- 17, April~1, 2017.

\end{thebibliography}
\newpage
\end{document}